\documentclass[letterpaper]{JHEP3}
\usepackage[xdvi]{graphicx} 
 
\title{Plane Waves: To infinity and beyond!}
\author{ Donald Marolf\footnote{E-mail: {\tt marolf@physics.syr.edu}}\ and
Simon F. Ross\footnote{E-mail: {\tt S.F.Ross@durham.ac.uk}}\\
$^*$ Physics Department, Syracuse University, Syracuse, New York
13244 USA\\
$^\dagger$ Centre for Particle Theory, Department of Mathematical
Sciences, University of Durham, South Road, Durham DH1 3LE UK}

\date{August, 2002}

\abstract{ We describe the asymptotic boundary of the general
homogeneous plane wave 
spacetime, 
using a construction of the `points
at infinity' from the causal structure of the spacetime as introduced by
Geroch, Kronheimer and Penrose. We show that this construction agrees
with the conformal boundary obtained by Berenstein and Nastase for the
maximally supersymmetric ten-dimensional plane wave.  We see in detail
how the possibility to go beyond (or around) infinity arises
from the structure of light cones.  We also discuss
the extension of the construction to time-dependent plane wave
solutions, focusing on the examples obtained from the Penrose limit of
D$p$-branes.
} 
\keywords{Plane Waves, PP-waves, causal structure, conformal
boundary} \preprint{SUGP-02/7-3, DCTP-02/63}

\begin{document}

\section{Introduction}

The plane waves of interest here 
are those particular solutions~\cite{Brinkmann,BPR,JH,AP,EK,exact} 
to theories incorporating Einstein-Hilbert gravity
for which the metric takes the form
\begin{equation}
\label{genplane}
ds^2 = - 2 dx^+ dx^- - \mu^2_{ij}x^i x^j (dx^+)^2 + dx^i dx^i, 
\end{equation}
where $\mu^2_{ij}$ is in general a function of $x^+$. For several reasons, 
these spacetimes have received considerable attention 
in the context of string
theory. Firstly, plane waves are
exact backgrounds for string theory, corresponding to exact conformal
field theories~\cite{Amati}. Secondly, the superstring action on certain
plane wave backgrounds in light cone gauge is purely
quadratic~\cite{Metsaev}, so the theory can be explicitly quantised.

Finally, an old result of Penrose~\cite{Plimit} that the spacetime
near any null geodesic is a plane wave has been used to relate string
theory on the maximally supersymmetric ten-dimensional plane wave to
four-dimensional field theory~\cite{BMN}. A null geodesic in $AdS_5
\times S^5$ which rotates on the $S^5$ is considered. Taking the
Penrose limit near this geodesic yields the maximally supersymmetric
ten-dimensional plane wave found by Blau, Figueroa-O'Farrill, Hull and
Papadopoulos (BFHP)~\cite{BFHP}, whose metric is 
\begin{equation}
\label{msplane}
ds^2 = - 2 dx^+ dx^- - \mu^2 x^i x^i (dx^+)^2 + dx^i dx^i, 
\end{equation}
$i=1,\ldots,8$. In the field theory dual to $AdS_5 \times S^5$, this
limit can be identified as a restriction to a subsector of `near-BPS'
operators of large R charge.

As in the usual AdS/CFT correspondence, an important step in
understanding the relation between string theory on the plane wave and
the field theory is to understand the asymptotic structure of the
bulk spacetime.
Berenstein and Nastase~\cite{BN} studied the asymptotic
structure of the BFHP plane wave (\ref{msplane}), constructing the
Penrose diagram by 
identifying a conformal transformation that takes this spacetime to
a measure one subset of the Einstein static universe 
(building in part on work of~\cite{KP}). 
They found that the conformal boundary consists of a (one-dimensional) null
line which spirals around the Einstein static universe as shown in
figure~\ref{ESU}. This line acts as both past and future boundary for
the plane wave, 
so one may pass beyond (or around) infinity and return
to the original spacetime. 
In analogy with the usual picture of the
CFT dual to an asymptotically AdS space,
Berenstein and Nastase suggested that the plane wave
possesses a holographic dual description in terms of a quantum
mechanics on this boundary.

\FIGURE{
    \includegraphics[width=0.3\textwidth,height=0.4\textheight]{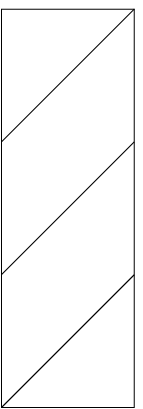}
\caption{A 1+1 dimensional slice of the
Einstein static universe along a great circle of the sphere.  Left
and right sides of the diagram are identified.  The winding null ray
indicates the conformal boundary of the original BFHP
plane under the Berenstein-Nastase conformal embedding \cite{BN}.}\label{ESU}}

In this paper, we consider the asymptotic
structure of the more general plane wave metric (\ref{genplane}). We
will be able to treat the case where $\mu_{ij}^2$ is constant in full
generality. Specific examples of interest in string theory include the
maximally supersymmetric eleven-dimensional solution~\cite{11d}
obtained from the Penrose limit of $AdS_4 \times S^7$ and $AdS_7
\times S^4$~\cite{ads4}, and partially supersymmetric plane waves in
ten dimensions~\cite{Cvetic,Bena,Michelson}---a notable example in the latter
category is the Penrose limit of the Pilch-Warner
flow~\cite{usc,leo,durham}. We will also give a discussion of the more
general case of non-constant $\mu_{ij}^2$, focusing on the plane waves
obtained from the Penrose limit of D$p$-branes~\cite{leo}.

Since these metrics are not conformally flat, it is difficult to apply
the conformal scaling approach used in~\cite{BN} to characterise the
asymptotic behaviour. Instead, we will use an approach introduced by
Geroch, Kronheimer and Penrose~\cite{Geroch}, which characterises
points at infinity (dubbed `ideal points') purely in terms of the
causal structure of the spacetime. This approach has the advantage of
being applicable to a wider class of 
spacetimes and 
gives an obvious physical interpretation to the points at infinity.

We begin by reviewing the technique and 
discussing its
application to the BFHP case (\ref{msplane}) in
section~\ref{allplus}. This will allow us to exhibit the close
relationship between this method and the conformal boundary method
used by Berenstein and 
Nastase and in addition to see what further insight is gained.
The set of ideal points for this spacetime is precisely the
conformal boundary of Berenstein and Nastase.

We then go on to study the general case of constant $\mu_{ij}^2$ in
section~\ref{gen}. So long as $\mu_{ij}^2$ has at least one positive
eigenvalue, the construction is very similar to the previous case, and
we find that the set of ideal points of the spacetime forms a
one-dimensional null line. We briefly comment on the rather unphysical
case of all negative eigenvalues.

There are also interesting examples with non-constant $\mu_{ij}^2$;
for instance, the Penrose limits of near D$p$-brane
spacetimes~\cite{leo} which should have dual descriptions based on
the gravity/gauge-theory dualities of~\cite{IMSY}. It is more
difficult to construct the ideal points in these cases, and we have
not attempted to do so for arbitrary values of the functions
$\mu_{ij}^2(x^+)$. In section~\ref{Dp}, we  
discuss the construction for the specific
examples considered in~\cite{leo}.  We find
that these examples also have one-dimensional null line segments of ideal
points. In the case $p=4$, the past and future points are identified,
but not for $p=0,1,2$. We conclude with a brief discussion of the
possible physical implications of our results and interesting future
directions in section \ref{disc}.

\section{The  BFHP plane wave}
\label{allplus}

In this section, we will study the asymptotic structure of the BFHP
plane wave (\ref{msplane}) using the method of Geroch, Kronheimer and
Penrose \cite{Geroch}, to set the stage for our later discussion of
more general plane waves. 

Let us first review the method. Our aim, given a spacetime manifold
$M$, is to construct a suitable completion $\bar M$ by adjoining some
suitable ideal points to $M$. We begin by considering the causal
structure of $M$, as encoded in the past-sets of $M$.\footnote{This
approach to determining the appropriate completion assumes that the
spacetime $M$ is strongly causal.} We define an IP (indecomposable
past-set) to be the causal past $I^-[\gamma]$ of any timelike curve
$\gamma$ in the spacetime. Note that $\gamma$ need not be complete.
Similarly, the causal future $I^+[\gamma]$ is called an IF. An IP such
that $I^-[\gamma] = I^-(p)$ for some point $p$ is called a proper IP,
or PIP. In this case, $p$ is the future endpoint of $\gamma$. If
alternatively $\gamma$ is future-endless, $I^-[\gamma]$ is a terminal
IP, or TIP. PIFs and TIFs are similarly defined.

The idea is to construct the ideal points from the TIPs and TIFs; that
is, we identify the additional points in terms of the spacetime
regions they can physically influence or be influenced by. Put another
way, we will add endpoints to endless curves, in such a way that two
curves go to the same point if they have the same past or future. Of
course, it is not sufficient to simply adjoin all the TIPs and all the
TIFs to the spacetime: some ideal points (such as the timelike
singularity of Reissner-Nordstr\"om or the conformal boundary of AdS)
naturally have curves approaching them from both the future and the
past. We must therefore identify some TIPs and TIFs. 

The construction proceeds by considering the set $\hat M$ of IPs and
the set $\check M$ of IFs of $M$. There are natural maps from $M$ to
each of these spaces, sending a point $p \in M$ to the PIP (PIF) $P =
I^-(p)$ ($P^* = I^+(p)$), so each of these sets gives a partial
completion of $M$. Next, consider $M^\sharp$, formed by taking the
union $\hat{M} \cup \check{M}$ and identifying all proper IPs $P$ with the
corresponding  proper IFs $P^*$. The proposed completion $\bar{M}$ is then
defined by giving an appropriate topology to $M^\sharp$ and taking a
quotient by the minimal set of identifications $R_H$ necessary to make
the space Hausdorff, $\bar{M} = M^\sharp/R_H$. 

Thus, to construct the points at infinity for the plane wave metric,
we will need to find the different TIPs and TIFs of the spacetime, the
pasts and futures of inextensible timelike curves, and determine which
TIPs and TIFs we must identify. The latter part is the more subtle
issue; we will not carry out the full topological program\footnote{In
fact, a modification of the procedure for constructing the
identifications given in \cite{Geroch} is necessary. The details of
our preferred method of constructing identifications will be described
in a forthcoming work.}, but will simply argue that if there is a set
of points $p_n \in M$ such that the PIPs $P_n$ approach some TIP $T$
and the PIFs $P_n^*$ approach some TIF $T^*$, then $T$ and $T^*$ must
represent the same ideal point.

\subsection{TIPs and TIFs for the BFHP plane wave}
\label{apply}

We first consider the BFHP plane wave using the Einstein static
universe coordinates introduced in~\cite{BN}. This will make it
obvious that the set of ideal points is precisely the one-dimensional
conformal boundary of the spacetime. We will then reproduce this
result in the original coordinates (\ref{msplane}), to enable
generalisation to other plane waves.

In~\cite{BN}, a coordinate transformation was found which brings the
metric (\ref{msplane}) to the form 
\begin{equation}
ds^2 = {1 \over 4 |e^{i \psi} - \cos \alpha e^{i\beta}|^2} (-d\psi^2 +
d\alpha^2 + \cos^2 \alpha d\beta^2 + \sin^2 \alpha d\Omega_7^2).
\end{equation}
The metric in parentheses is the metric of the Einstein static
universe (ESU) $R \times S^9$. Since the causal structure is invariant
under conformal transformations, the IPs \& IFs of the plane wave are
in one-to-one correspondence with the IPs \& IFs of the ESU. In the
ESU, all of the IPs are PIPs with the exception of an IP which is the
whole spacetime (realised, for example, as the past of the curve
$\alpha=\beta=0$) which is a TIP, corresponding to an ideal point
$i^+$ at future timelike infinity. The whole spacetime is also a TIF,
corresponding to the ideal point $i^-$.

The conformal factor relating the ESU to the plane wave diverges on
the points $\alpha=0$, $\psi=\beta$. This is a one-dimensional null
line, which orbits on a great circle of the $S^9$; we call it ${\cal
I}$. The conformal boundary of the plane wave is ${\cal I}$, together
with the points at infinity $i^\pm$ in the ESU.

{}From the point of view of the study of IPs, the effect of this
conformal rescaling is that some of the PIPs of the ESU become TIPs in
the plane wave. More formally, future-endless timelike curves in the
plane wave spacetime 
can be mapped to curves in the ESU which
either have an endpoint on one of the points $p
\in {\cal I}$ or extend to $i^+$. Thus, the TIPs of the plane
wave are identified with the set ${\cal I}$, together with the point
$i^+$. Similarly, the set of TIFs is ${\cal I}$ together with the
point $i^-$. Clearly, the TIP $T$ and the TIF $T^*$ associated with
the same point on $\cal I$ must be identified; to use our general line
of argument, note that if a sequence of points $p_n$ define PIPs $P_n$
which approach $T$, the points $p_n$ must approach a point on $\cal I$
in the ESU, and hence the PIFs $P_n^*$ will approach $T^*$, forcing
the identification. Thus, the set of ideal points is the same as the
conformal boundary.

We will now show how this result emerges in the original coordinate
system (\ref{msplane}). This detailed discussion sets the stage for
the next section, but it is also intrinsically useful: it is unusual
for a null conformal boundary to have non-trivial regions of the
spacetime both to the future and the past, and the discussion will
help us to understand how this happens. The possibility of the
boundary both influencing and being influenced by the bulk seems
important for holography, so we should carefully understand the
identifications between the past and future boundaries.

It is still simple to identify the possible TIPs in this
coordinate system. Consider any causal curve $\gamma$.  If $\gamma$
reaches arbitrarily large values of $x^+$, it will become clear below that
its TIP is  the
entire spacetime.  We thus turn to the remaining case where $\gamma$
asymptotes in the future to a constant value of $x^+$.  No other
behaviour is allowed as $x^+$ must increase along future directed
causal curves\footnote{This follows from the fact that
(\ref{msplane}) is the Minkowski metric plus a term that is second
order in $dx^+/d\lambda$ for any parameter $\lambda$.}.  Now $x^-$
must diverge as a curve asymptotes to a constant value of $x^+$.  To
see this, note first that some coordinate must diverge or the curve
would have an endpoint in the spacetime.  If this coordinate is not
$x^-$, then some $x^i$ and thus $\ln x^i$ and $d(\ln x^i)/dx^+$ must
diverge.  But this requires $\frac{dx^i}{dx^+} \gg x^i,$ and any causal
curve which asymptotes to $x^+$ with diverging $x^i$ satisfies
\begin{equation}
2 \frac{dx^-}{dx^+}  \ge \left( \frac{dx}{dx^+}\right)^2  - \mu^2 x^2 \sim  
\left( \frac{dx}{dx^+}\right)^2.
\end{equation}
As a result, $x^-$ must diverge even faster than $x^i$.

Let us now write the metric (\ref{msplane}) in the form
\begin{equation}
ds^2 = ds_0^2 + \Delta ds^2,
\end{equation}
with
\begin{eqnarray}
ds_0^2 &=& -2 dx^+ dx^- + dx^2,  \cr
\Delta ds^2 &=& - \mu^2 x^2 (dx^+)^2.
\end{eqnarray}
It is 
then
clear that any pair of coordinates
$(x^+_1,x^-_1,x_1),(x^+_2,x^-_2,x_2)$ which is causally connected in
Minkowski space is also causally connected in the BFHP plane wave.  In
(conformally compactified) Minkowski space, the past of a point with
fixed $x^+$ and divergent $x^-$ contains all points with $x^+ \le
x^+_1$.  This immediately tells us that any curve which asymptotes to
$x^+ = x^+_1$ contains all points with $x^+ \le x^+_1$ in its TIP.
But since $x^+$ is a non-decreasing function along any (other) causal
curve, no point with $x^+ > x^+_1$ can be in the past of our curve.
Thus the TIP of our curve is exactly the set $\{x^\mu : x^+ \le x^+_1
\}$.  The future boundary is thus a one dimensional line parametrised
by $x^+$. Similarly, the past boundary is again a one dimensional line
parametrised by $x^+$.

Now we wish to understand the identification between TIPs and
TIFs. It is easy to show that the curves
\begin{equation}
\label{c1}
x^i = A^i \sin(\mu x^+), \ \ 
x^- = \frac{1}{2}x^i \partial_+ x^i  + C(x^+)
\end{equation}
are null (in fact, they are null geodesics) when $C(x^+)$ is constant
and that they are either timelike or null so long as $C(x^+)$ is an
increasing function of $x^+$. In more detail, we may write
\begin{equation}
\label{xm1}
x^- = \frac{1}{2} A^2 \mu \sin(\mu x^+) \cos(\mu x^+) + C(x^+)
= \frac{1}{2}\mu x^2 \cot(\mu x^+) + C(x^+).
\end{equation} 
We thus learn that the future of the origin ($x=0,y=0,x^+=0,x^-=0$)
contains all points with
\begin{equation}
\label{ineq}
x^- \ge  \frac{1}{2}\mu x^2 \cot(\mu x^+).
\end{equation}
In fact, the curves (\ref{c1}) form the null cone leaving the origin.
As a result, any point in the future of the origin satisfies
(\ref{ineq}) until our family of null curves caustics at $x^+ =
\pi/\mu$. Since (\ref{ineq}) is satisfied by any $x^-,x\neq 0$ for
$x^+=\pi/\mu$, we see immediately that the origin is in the past of
almost every point on the $x^+ = \pi/\mu$ plane.  But the future of an
event is a closed set, so the origin must lie in the past of all
points with $x^+ \ge \pi/\mu$. Since we can use the symmetries to
translate the origin to an arbitrary point in the surface $x^+=0$, it
follows that the past of any point with $x^+ \ge \pi/\mu$ contains the
whole region $x^+ < 0$. It is now clear that the TIP of any curve
reaching arbitrarily large values of $x^+$ is the entire spacetime.

\FIGURE{\includegraphics[width=0.4\textwidth]{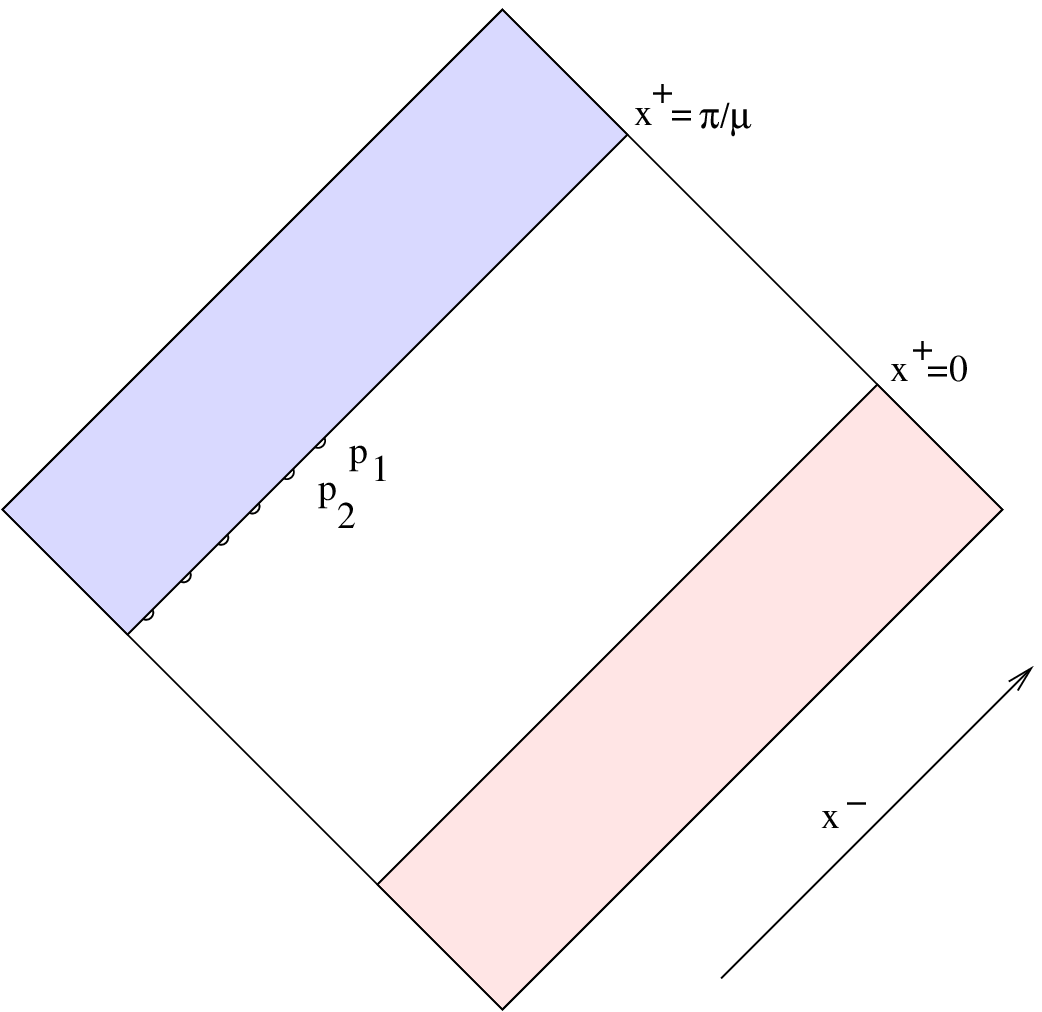}
  \caption{The points $p_n$, the TIF $\{x^\mu: x^+ \geq \pi/\mu \}$ and
  the TIP $\{x^\mu : x^+ \le 0\}$ in the $x^\pm$ plane.}
\label{CausRel}}

A sequence of points $p_n$ with $x^+(p_n)=\pi/\mu$ and
$x^-(p_n)\rightarrow - \infty$ define PIFs $P_n^* = I^+(p_n)$ which
approach the TIF $T^* = \{x^\mu : x^+ \geq \pi/\mu \}$. Using the
result of the previous paragraph, the PIPs $P_n = I^-(p_n)$ defined by
this sequence each contain the TIP $T= \{x^\mu : x^+ \le 0\}$. Using
an appropriately time-translated version of (\ref{ineq}), in the limit
$x^-(p_n)\rightarrow - \infty$, the past of $p_n$ contains only points
with $x^+ \leq 0$. Thus, the $P_n$ approach $T$. Hence, we should
identify $T$ and $T^*$. By time-translation, every TIF $\{x^\mu : x^+
\geq x^+_1 + \pi/\mu \}$ is identified with the TIP $\{x^\mu : x^+
\leq x^+_1\}$.  This completes the demonstration of the identification
between TIPs and TIFs.  Again we see that it is possible to
pass beyond infinity and return to the original spacetime.

We can also use this to show that the line of TIPs is a null line in
$\hat M$. There is a natural causal relationship between the IPs in
$\hat M$, under which $P$ is to the causal past of $Q$ in $\hat M$ if
$P \subset Q$. Under this causal relationship, the TIPs form a causal
line; the TIP $P$ defined by $x^+ \leq x^+_P$ is clearly to the past
of the TIP $Q$ defined by $x^+ \leq x^+_Q$ iff $x^+_P \leq x^+_Q$. Now
two causally-separated points $P$ and $Q$ are null separated if there
is a sequence of points $P_n$ which approaches $P$ but none of the
points $P_n$ is causally separated from $Q$. Consider $x^+_Q = x^+_P +
\epsilon$ for some small $\epsilon$. We have just demonstrated that
the sequence of PIPs $I^-(p_n)$ where $x^+(p_n) = x^+_P + \pi/\mu$
approaches $P$, but $I^-(p_n)$ contains points with $x^+ = x^+_P +
\pi/\mu - \delta$, which are not in $Q$. Hence 
the
$P_n$ are not causally
related to $Q$, and $P$ and $Q$ must be null separated.\footnote{Note
that points $P$ and $Q$ with $x^+_Q > x^+_P + \pi/\mu$ are timelike
separated; this expresses the fact that the null boundary is `winding
around' in $\hat M$, as we saw explicitly in the conformal rescaling
construction.}

As advertised, we have reconstructed the result that the set of ideal
points is the null line $\cal I$ parametrised by $x^+$ together with
the two points $i^\pm$.  We will see below that this construction
generalises readily to many other cases.

\section{Homogeneous plane waves}
\label{gen}

For homogeneous plane waves, where the coefficient functions
$\mu^2_{ij}$ in (\ref{genplane}) are constants, the non-trivial
components of the Weyl tensor are
\begin{equation}
C_{+i+j} = \mu^2_{ij} - {1 \over n} \mu^2_{kk} \delta_{ij},
\end{equation}
$i,j,k = 1, \ldots, n$. Thus, the metric is conformally flat iff
$\mu^2_{ij} = \mu^2 \delta_{ij}$, that is, it is conformally flat only
for the maximally symmetric case considered above (and a case with
$\mu^2$ negative discussed below). There are many examples of
particular interest to string theory which do not fall into this
category; and for which it is therefore quite complicated to find an
explicit Penrose-style conformal compactification.  For instance, the
maximally supersymmetric 11d plane wave~\cite{11d}, the Penrose limit
of $AdS_3 \times S^3 \times M_4$~\cite{Blau},\footnote{The
six-dimensional geometry obtained by the Penrose limit of $AdS_3
\times S^3$ is conformally flat, but on crossing it with a
four-dimensional flat space, we lose this property.} and Penrose
limits of less supersymmetric ten-dimensional geometries, such as the
Pilch-Warner solution~\cite{usc,leo,durham}, are not conformally flat.
(Other examples where different homogeneous plane waves arise as
Penrose limits are discussed in~\cite{Hubeny,Oz,Ali}).  Nevertheless,
one can readily study the infinity of these spacetimes using the
methods \cite{Geroch} just described.  In this section, we will find
the TIPs and TIFs for all homogeneous plane waves.

For almost all cases, the set of ideal points is identical to that
found in the previous section, namely a single null line. This
suggests that these more general examples might also have
quantum-mechanical dual descriptions. This is the more surprising as
the discussion includes cases where some of the eigenvalues of
$\mu^2_{ij}$ are negative. The presence of such negative eigenvalues
is believed to lead to an instability in the string sigma-model, as
was discussed in \cite{leo}. 
Nevertheless, from the point of view of the asymptotic
structure of spacetime, these examples behave in the same way as cases
with all eigenvalues positive.

We consider a spacetime of the form (\ref{genplane}). For arbitrary
$\mu^2_{ij}$, we can make a rotation on the $x^i$ to make $\mu^2_{ij}$
diagonal. We assume there is at least one positive eigenvalue, and
write the metric as
\begin{equation}
\label{mixedmetric}
ds^2 = - 2 dx^+ dx^- - (\mu_1^2 x_1^2 + \ldots + \mu_j^2 x_j^2 - m_1^2
y_1^2 - \ldots - m_{n-j}^2 y_{n-j}^2) (dx^+)^2 + dx^i dx^i + dy^a dy^a,
\end{equation}
where $\mu_1 \geq \mu_2 \geq \ldots \geq \mu_j$. As in the previous
discussion, the first step is to determine the TIPs and TIFs. As
before, a future-directed timelike curve $\gamma$ must move in the
direction of increasing $x^+$, so it will either extend to arbitrary
values of $x^+$ or approach some constant value of $x^+$.  Again,
it will become clear below that a
causal curve reaching arbitrarily large values of $x^+$ has the entire
spacetime as its TIP.  All other causal curves asymptote in the future
to a constant value of $x^+$.  Curves that asymptote to distinct
values of $x^+$ have distinct TIPs, while we again claim that all
curves asymptoting to the same value of $x^+$ have the same TIP.  If
$\mu^2_{ij}$ has all positive eigenvalues, the proof of this statement
is as before: the curve will reach divergent $x^-$, and the past of
any point is at least its past in the Minkowski metric. If some of the
eigenvalues are negative, the statement is still true, but the proof
is slightly more complicated than in the positive definite case and
has been placed in appendix A.  In particular, this appendix shows
that the TIP of any causal curve asymptoting to $x^+=x^+_1$ consists
exactly of those points with values of $x^+$ smaller than the
$x_1^+$. Thus, the TIPs form a one dimensional line parametrised by
$x^+$, together with the single point $i^+$ (which one may regard as
the continuation of the line to $x^+=+\infty$).  Similarly, the TIFs
form a line parametrised by $x^+$, together with $i^-$ (at
$x^+=-\infty$).

Finally, 
we need to determine the relation between TIPs and TIFs. The following curves are causal
so long as $C(x^+)$ is an increasing function and $A^i,B^i$ are constants:
\begin{eqnarray}
x^i &=& A^i \sin(\mu_i x^+), \cr
y^a &=& B^a \sinh(m_a x^+), \cr
x^- &=& \frac{1}{2}(x^i \partial_+ x^i  + y^i \partial_+ y^i) + C(x^+)  \cr
&=& \frac{1}{2}
\frac{(A^i)2}{2} \mu_i \sin(\mu_i x^+) \cos(\mu_i x^+) + \frac{(B^a)^2}{2} 
m_a \sinh(m_a x^+) \cosh(m_a x^+) + C(x^+) \cr
&=& \frac{\mu_1 x_1^2}{2} \cot(\mu_1 x^+) + \ldots
\end{eqnarray}
Thus, the future of the origin ($x=0,y=0,x^+=0,x^-=0$) is exactly the set
with 
\begin{equation}
\label{ineq2}
2x^- \ge \mu_1 x_1^2 \cot(\mu_1 x^+) + \ldots
\end{equation} 
until our family of null curves encounters the first caustic at $x^+ = \pi/\mu_1$.

Now, take $x^+ = \pi/\mu_1 - \epsilon.$ By taking $\epsilon$
sufficiently small, we can arrange for (\ref{ineq2}) to be satisfied
for any $x_1\neq 0$ (since, from the convention that $\mu_1$ is the
largest eigenvalue, the terms not written explicitly will remain
finite\footnote{Except for eigenvalues degenerate with $\mu_1$, which
contribute in exactly the same way as the term shown.}  as $\epsilon
\to 0$).  Since the future is a closed set, we see that the origin is
in the past of every point on the $x^+ = \pi/\mu_1$ plane.  By
translation invariance, the past of any point with $x^+=x^+_1$
contains all points with $x^+ \le x^+_1 - \pi/\mu_1$.  In particular, 
it is now clear that the TIP of any curve reaching arbitrarily
large values of $x^+$ is the entire spacetime. The remaining
step of identifying future boundary points with past boundary points
now follows just as in section \ref{allplus} by arguing that the
sequence of points $p_n$ such that $x^+=x^+_1$ and $x^- \to -\infty$
has PIFs $I^+(p_n)$ approaching the TIF $\{ x^\mu : x^+ \geq x^+_1 \}$
and by (\ref{ineq2}) has PIPs $I^-(p_n)$ approaching the TIP $\{ x^\mu
: x^+ \leq x^+_1-\pi/\mu_1\}$. Thus, this TIP and TIF are identified
for each $x_1^+$. We can show as before that this is a null line.

To summarise, we have shown that for constant $\mu^2$ with at least
one positive eigenvalue, the set of ideal points is a single null line
parametrised by $x^+$, together with the two points $i^\pm$. 

\subsection{No positive eigenvalues}

For completeness, we now mention the case of the plane wave with no
positive eigenvalues. Leaving aside flat space ($\mu^2_{ij}=0$), these
cases with negative eigenvalues are unphysical, as no such solution
can arise in string theory (so long as we keep the symmetry associated
with the Killing vector $\partial_+$, so the dilaton is constant). The
point is that the non-vanishing Ricci tensor component, $R_{++}$, is
essentially just the sum of the eigenvalues.  However, negative
$R_{++}$ in the Einstein frame violates the weak null energy
condition.  Hence, these backgrounds are not solutions to any known
string theory.

Nevertheless, we will consider this case briefly, because it results
in a different structure than the cases with positive eigenvalues. For
simplicity, we will consider the special case where the metric is
conformally flat, and take $\mu^2_{ij} = - \delta_{ij}$. The metric is
then
\begin{equation}
\label{allminus}
ds^2 = -2 dx^+ dx^- +  y^2 (dx^+)^2 + dy^2.
\end{equation}
We can use the conformal boundary technique; simply apply the change
of variables
\begin{eqnarray}
v &=& x^- - (y^2/2) \tanh x^+ \cr
u &=& \tanh x^+ \cr
\tilde y &=& y/\cosh x^+,
\end{eqnarray}
to obtain the metric
\begin{equation}
ds^2 = (1-u^2)^{-1} \left( -2du dv + d\tilde y^2 \right),
\end{equation}
with $-1 < u < 1.$  This shows explicitly that  the spacetime
(\ref{allminus}) is conformally equivalent to a slice of Minkowski
space bounded by two null planes.  

As a result, the conformal boundary of (\ref{allminus}) consists of
null planes at $u=\pm 1$ and two null lines which represent the
intersection of the region $-1 < u <1$ with the past and future
infinity of the Minkowski space described by the conformally rescaled
metric $d\tilde s^2 = -2du dv + d\tilde y^2.$ We see that this is
qualitatively different from the other homogeneous plane waves.

\section{General plane waves}
\label{Dp}

We will now discuss the cases where $\mu^2_{ij}$ are functions of
$x^+$. A particularly interesting example of this type is provided by
the Penrose limits of near D$p$-brane spacetimes obtained
in~\cite{leo}. Since the near D$p$-brane geometries for $p \le 4$ all
have dual field theory descriptions~\cite{IMSY}, these cases may also
lead to interesting insights into gauge/gravity connections. The
resulting plane waves can have eigenvalues which become negative for
some range of $x^+$.

When $\mu^2_{ij}$ is a function of $x^+$, there may be limitations on
the range of $x^+$ (if the functions diverge at some finite $x^+$),
and this can lead to interesting structures of TIPs associated with
timelike curves which approach the maximum value. However, this
structure will depend in a complicated way on the particular functions
involved, and we are more interested in considering the effects on the
one-dimensional line of ideal points parametrised by $x^+$ we
previously considered. Therefore we will focus on the TIPs associated
with causal curves which asymptotically approach some constant value
of $x^+$ where $\mu^2_{ij}$ is not diverging. It is easy to extend the
previous arguments to show that the TIP associated with such curves is
again a region of the form $x^+ \leq x^+_1$, so that there is a set of
TIPs parametrised by $x^+$ (and similarly for TIFs). The details are
given in the appendix. To see whether these TIPs and TIFs are
identified, we need to consider specific examples, 
to which we now turn.

\subsection{Limits of D$p$-branes}

We want to consider the plane waves obtained from Penrose limits of
near D$p$-branes for $p \le 4$, as in \cite{leo}.  Null geodesics in the
near D$p$-brane spacetimes emerge from the brane singularity at a
finite affine parameter in the past and return at a finite affine
parameter in the future.  As a result, the associated plane waves are
singular with the eigenvalues of $\mu^2$ diverging at finite values of
$x^+$, say $x^+=\pm x^+_0$.  The divergence is of the form $\mu^2
\rightarrow + \infty$ for $p=4$ and $\mu^2 \rightarrow -\infty$ for
$p=0,1,2.$

Due to the singularities at $x^+ = \pm x^+_0$, it is clear that
infinity is formed only from the TIPs and TIFs of causal curves that
asymptote to null planes with $|x^+| < x^+_0$.  As a result, the
boundary will consist of the singular null planes together with a null
line (or perhaps two) at infinity.  It remains only to determine any
identifications between past and future infinity.

Let us begin with the case $p=4$.  Here, we see from equation (5.42)
of \cite{leo} that the eigenvalues are always positive and diverge as
$(x^+-x^+_0)^{-2}$ at the singularities.  Thus, the frequency of the
harmonic oscillator diverges as $(x^+-x^+_0)^{-1}$ and the oscillator
oscillates infinitely many times before reaching the singularity.
{}From our earlier discussion of the behaviour of light cones, it follows
that each point on future infinity is thus identified with some point
on past infinity, and vice versa.  This is true even for the points
close to the singularities.

Turning now to the cases $p=0,1,2,$ recall that the basic feature
associated with identifications between past and future infinity is
the existence of two values $x^+_1,x^+_2$ of $x^+$ such that the
future of any point on the plane $x^+=x^+_1$ contains every point on
the plane with $x^+=x^+_2$.  In particular, it contains points
$(x^-,x^i,x^+=x^+_2)$ with arbitrarily large and negative values of
$x^-$.  In section \ref{gen}, such behaviour was the a result of a
harmonic oscillator oscillating though half of a period.

Our strategy here is much the same as in the constant $\mu^2$ cases, though
the treatment is necessarily less explicit.  The reader may verify
that the null cone of a point is again traced out by null geodesics
satisfying
\begin{eqnarray}
\partial_+^2 x^i &=& -\mu^2_{ij} x^j, \\
x^- &=& \frac{1}{2} x^i \partial_+ x^i + C,
\end{eqnarray}
where $C$ is a constant.  Points on past and future infinity will be
identified only if there is some sequence of points on this null cone
for which $x^-$ diverges while $x^i$ remains finite. It is clear that
this in turn can happen only if the dimensionless ratio $x^i
\partial_+ x^i /x^2$ diverges toward negative infinity.  Consulting
\cite{leo}, we see that for any $x^+$ the coordinate called $\phi$ in
that reference is always associated with the largest eigenvalue of
$\mu^2$, and thus the shortest period.  As a result, it is sufficient
to consider only this direction and to set the other coordinates to
zero.

Because of the complicated dependence on $x^+$, the above expressions
are difficult to analyse analytically.  However, it is straightforward
to solve the equations numerically with various initial data $\phi(0),
\partial_+ \phi(0)$.  Of course, only the ratio $\frac{\partial_+
\phi(0)}{\phi(0)}$ can effect the scale invariant quantity $x^i
\partial_+ x^i /x^2 = \partial_+\phi / \phi$.  If identifications
occur, the ratio $x^i \partial_+ x^i /x^2 = \partial_+\phi / \phi$
evolves from positive infinity when the light cone is emitted to
negative infinity when the identification occurs.  In particular this
ratio, and thus $\partial_+ \phi$, goes through zero in between.  If
we wish to find identifications, it is to our advantage to choose a
light cone which spends as much time as possible near $x^+=0$ where
the harmonic oscillator frequencies are highest.  As a result, it is
best to choose the zero of $\partial_+\phi$ to occur at $x^+=0$.

Thus, one need only consider the initial data
$\phi(0)=1,\partial_+\phi(0)=0.$ In each case ($p=0,1,2$), one finds that
the ratio $\partial_+\phi/ \phi$ is in fact bounded below.  More
specifically, at $x^+=0$ we have $\phi=1$, after which $\phi$
decreases to a positive minimum and then turns around and begins to
increase.  The coordinate $\phi$ is still increasing when the
singularity is encountered.  We therefore conclude that there are no
identifications between past and future infinity in these spacetimes.

\section{Discussion}
\label{disc}

We have discussed the asymptotic structure of plane waves using the
approach to constructing ideal points based on past-sets as introduced
in~\cite{Geroch}. We first reconsidered the asymptotic structure of
the maximally-supersymmetric ten-dimensional plane wave from this
point of view. We showed that we can reproduce the results
of~\cite{BN}, who constructed the Penrose diagram for this case. We
believe the current approach offers some useful additional insight
into the relation between bulk and boundary in this case; in
particular, it emphasises the importance of the identifications
between past and future boundaries, which are quite non-trivial in the
original coordinate system.
As is apparent form the conformal embedding of the BFHP wave into
the Einstein static universe, these ideal points can also be
reached by spacelike curves.  While we have not addressed this
explicitly, it is clear from our analysis that any spacelike curve
which asymptotes to $x^+=x^+_0$ with sufficiently rapidly
diverging $x^-$ reaches the
same ideal point as do causal curves with the same asymptotics.

The real power of the approach of~\cite{Geroch}, however, is that it
can easily be extended to more general metrics. We have shown that the
general homogeneous plane 
wave has
an identical causal boundary,
with a one-dimensional line of ideal points, so long as
there is at least one positive eigenvalue of
$\mu^2_{ij}$. Surprisingly, the asymptotic structure is unchanged when
we introduce negative eigenvalues, even though these will produce
dramatic changes in the dynamical behaviour of both particles and
strings 
in
the spacetime.  So long as there is at least one
positive eigenvalue, there remains the possibility
to pass beyond infinity and return to the original spacetime.

The most interesting direction for future work is to explore the
consequences for possible holographic dual descriptions of these
spacetimes. As in the previously-studied maximally supersymmetric
case~\cite{BN}, the structure of the asymptotic boundary, and in
particular the fact that it is both to the future and the past of
regions of spacetime, can be viewed as evidence in favour of the
existence of a quantum-mechanical dual description. For cases where a
dual field theory description is expected to exist, such as the
maximally supersymmetric 11d plane wave which arises in the Penrose
limit of $AdS_4 \times S^7$, this result is satisfying but perhaps
unsurprising. However, we find that the relevant behaviour is far more
general than this. It would be especially interesting if further
evidence could be found for the existence of a holographic dual for
the cases with some negative eigenvalues.

In the more general case where $\mu^2_{ij}$ are functions of $x^+$,
the analysis is more complicated, but in many cases at least some of
this structure will survive. 
We have explored in detail the asymptotic structure
of the cases arising from Penrose limits of D$p$-branes,
finding that there is a set of TIPs and TIFs which form a
one-dimensional line segment parametrised by $x^+$; the TIPs and TIFs
are identified for $p=4$ but not for $p=0,1,2$. It would be
interesting to extend this further to investigate more time-dependent
examples.

\FIGURE{\includegraphics[width=0.4\textwidth]{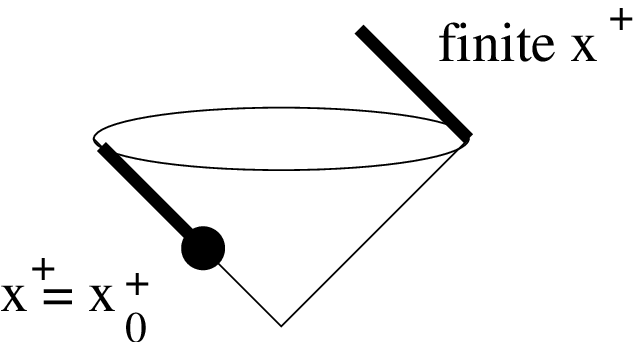}\caption{The
asymptotic
structure of a plane wave that has $\mu^2_{ij}=0$ before $x^+=x^+_0.$ The
null plane $x^+=x^+_0$ intersects past infinity only at the marked point.
Heavy line segments indicate potential identifications between
past and future infinity.}
\label{asMink}}

One interesting 
case occurs when 
the matrix $\mu_{ij}^2$ remains diagonal
but vanishes in, say, the asymptotic past.  Suppose in particular that
no harmonic oscillator oscillates more than some finite number of
times in the past.  In this case past infinity is essentially that of
flat Minkowski space!  The details become clear if we suppose that
$\mu_{ij}^2$ is strictly zero before some $x^+ = x^+_0$.  
The region to the
past of $x^+_0$ is the part of Minkowski space below some null plane
$P$, whose past boundary is that of Minkowski space except for a half
line to the future of the point on past infinity where $x^+=x^+_0$
(see figure \ref{asMink}). 
Thus, whether and how this half line is identified with part
of future infinity is determined by the behaviour of the harmonic
oscillators to the future of $x^+_0$.

Another avenue for future development is to explore the
asymptotic structure for pp-waves, where a more general dependence on
$x^i$ is allowed. These are also exact backgrounds for string theory,
at least to all orders of perturbations in $\alpha'$. Specific
examples have recently been considered in~\cite{maoz}, where string
propagation on such backgrounds was considered, and connections to
two-dimensional integrable systems were noted.
Finally, we plan to address detailed issues of the identifications that were
suppressed here in a forthcoming paper.

\appendix

\section{Appendix: TIPs of the general plane wave}

In this appendix we show that the TIP of any causal curve asymptoting
in the future to $x^+ = x^+_1$ is given by $\{ x^\mu : x^+ \le x^+_1
\}$. In particular, all causal curves which asymptote to the same
value of $x^+$ have the same TIP. We will consider the general metric
(\ref{genplane}), assuming only that the functions $\mu^2_{ij}(x^+)$
are regular in an open neighbourhood of $x^+ = x^+_1$. 

The non-trivial step in showing that the TIP is the region $\{ x^\mu :
x^+ \le x^+_1 \}$ is to show that the past of the curve includes all
points on the surface $x^+ = x^+_1 - \delta$ for arbitrarily small
$\delta$. The past of this null surface will clearly include all
points with smaller $x^+$, so if we can show this, it will follow that
the TIP contains the open set $\{ x^\mu : x^+ < x^+_1 \}$. We also
know that the TIP cannot contain any points with $x^+ > x^+_1$, since
no causal curve can move in the direction of decreasing $x^+$. Hence,
since the TIP is a closed set, it will be precisely $\{ x^\mu : x^+
\le x^+_1 \}$.

We therefore need only consider the behaviour in a small neighbourhood
of the surface $x^+ = x^+_1$,
so
we may approximate the functions
$\mu^2_{ij}(x^+)$ by constants $\mu^2_{ij}(x^+_1)$ and diagonalise
this matrix in the neighbourhood of $x^+ = x^+_1$ by a rotation on the
$x^i$.  The result is a metric of the form (\ref{mixedmetric}). We
will also redefine the $x^+$ coordinate so that the surface to which
the curve asymptotes is at $x^+=0$.

Let us now study the causal future of the origin of coordinates in the
metric (\ref{mixedmetric}). Consider the new set of null curves given
by
\begin{equation}
x^i =  A_i \sinh (\mu_i x^+), \ \ 
y^a = B_a \sin (m_a x^+), \ \ 
x^- = (1/2) (\mu_i^2 A_i^2 + m_a^2 B_a^2) x^+.
\end{equation}
(Note that these are different from the causal curves considered
earlier; in particular, $\sin$ and $\sinh$ are interchanged.)  These
show us that all points satisfying
\begin{equation}
x^- \ge \frac{1}{2} x^+ \left( \sum_i \left( \frac{\mu_j x^j}{\sinh(\mu_j
 x^+)}  \right)^2 + \sum_b \left( \frac{m_b y_b}{\sin(m_b x^+)} \right)^2
\right)
\end{equation}
are in the causal future of the origin.  But for $x^+ < \delta_1$
and small enough $\delta_1$, the Taylor expansions yield
\begin{eqnarray}
x^- &\ge& \frac{x^2 + y^2}{2 x^+ (1- \epsilon_1)},  \ \ \ {\rm or} \cr
y^2 + x^2 &\le& 2(1-\epsilon_1) x^- x^+ 
\end{eqnarray}
for any $\epsilon_1$. More generally, for $\Delta x^+ < \delta_1$ and
small enough $\delta_1$, the future of the point $x^i=0,y^i=0,x^-=0,
x^+ = x^+_0$ contains all points satisfying
\begin{equation}
x^2 + y^2 \le 2(1-\epsilon_1) x^- \Delta x^+.
\end{equation}
Let us call this region $R_{x^+_0}$.

Now consider any causal curve that asymptotes in the future to the
surface $x^+=0$.  We want to show that this enters the future of the
point $x^i=0,y^i=0,x^-=0, x^+ = x^+_0<0$ for small enough $|x^+_0|$.
Explicitly, this future region $R_{x^+_0}$ is
\begin{equation}
y^2 + x^2 \le 2(1-\epsilon_1) x^- |x^+_0|.
\end{equation}
Since any point can be brought to $x^i=0,y^a=0,x^-=0, x^+ = x^+_0$ by
a symmetry transformation that does not act on $x^+$, we will have
shown that any causal curve which asymptotes to $x^+=0$ enters the
future of any point having $x^+ < 0$.

Consider then an arbitrary causal curve that asymptotes to $x^+=0$.
Below, we use dots to denote $d/dx^+$.  Causality implies
\begin{equation}
2 \dot{x}^- \ge (m_a^2 y_a^2 - \mu_i^2 x_i^2) + \dot{y_a}^2 + \dot{x_i}^2 
\ge (-\mu_i^2 x_i^2) + \dot{y_a}^2 + \dot{x_i}^2 .
\end{equation}
Now, some coordinate (and therefore some velocity) must diverge at
$x^+=0$.  If only $x^-$ diverges while $x_i,y_a$ remain bounded, then
it is clear that the curve enters $R_{x^+_0}$.  If $x_i$ do not
diverge, then some $y_a$ and $\dot{y_a}$ diverge, and close enough to $x^+=0$ we
achieve
\begin{equation}
2 \dot{x}^- \ge (1-\epsilon_2) (\dot{x_i}^2 + \dot{y_a}^2) .
\end{equation}
But if some $x_i$ diverge, then $\ln x_i$ and $\partial_+ \ln x_i$ diverge,
so that we have $\dot x_i \gg x_i$ and again we achieve $ 2 \dot{x}^- \ge
(1-\epsilon_2) (\dot{x_i}^2 + \dot{y_a}^2).$ We will use this relation
below, and also the corollary that $x^-$ must diverge at $x^+=0$.

For the final step, let us introduce a new (fiducial and completely
non-physical) metric
\begin{equation}
ds^2_{\rm fid} = -2 dx^+ dx^- + (1-\epsilon_2) (dx^2 + dy^2),
\end{equation}
and note that sufficiently close to $x^+=0$ our curve is also causal
with respect to this fiducial metric. So, if we choose any point
$x^i_1,y^a_1,x^-_1, x^+= -\epsilon_3$ close to $x^+=0$ on our causal
curve, the curve remains in the fiducial future light cone of this
point.  That is, the curve satisfies
\begin{equation}
y^2 + x^2 \le \frac{2}{1-\epsilon_2} (x^- - x^-_1) \epsilon_3.
\end{equation}
This controls the rate at which $x_i$ and $y_a$ can diverge with $x^-$.
For $2 \epsilon_3 \le (1-\epsilon_2) (1-\epsilon_1) |x^+_0|$, we see
that our causal curve cannot run to large $x^-$ without entering the
region
\begin{equation}
y^2 + x^2 \le 2(1-\epsilon_1) x^- |x^+_0|,
\end{equation}
and thus the future of $x_i=0,y_a=0,x^-=0,x^+=x^+_0$. 
Hence any casual curve $\gamma$ asymptoting 
to $x^+=0$ has this point, and thus any
point with $x^+<0$, in its causal past.  This completes the proof that the
TIP $I^-[\gamma]$ is $\{ x^\mu : x^+ \le 0 \}$.

\acknowledgments 
DM would like to thank Eric Gimon, Veronika Hubeny,
Finn Larsen, and Leo Pando-Zayas
for conversations that inspired this project.  DM was supported in
part by NSF grant PHY00-98747 and by funds from Syracuse
University. SFR was supported by an EPSRC Advanced Fellowship.

\end{document}